\documentclass{JINST}

\title{The Upgrade of the CMS RPC System during the First LHC Long Shutdown}

\author{M.~Tytgat$^a$\thanks{Corresponding author.}, A.~Marinov$^a$,
  P.~Verwilligen$^a$\thanks{Now at Universita e INFN, Sezione di Bari.}, N.~Zaganidis$^a$,
A.~Aleksandrov$^b$, V.~Genchev$^b$, P.~Iaydjiev$^b$, M.~Rodozov$^b$, M.~Shopova$^b$, G.~Sultanov$^b$,
Y.~Assran$^c$,
M.~Abbrescia$^d$, C.~Calabria$^d$, A.~Colaleo$^d$, G.~Iaselli$^d$,
F.~Loddo$^d$, M.~Maggi$^d$, G.~Pugliese$^d$,
L.~Benussi$^e$, S.~Bianco$^e$, M.~Caponero$^e$, S.~Colafranceschi$^{e,i}$, 
F.~Felli$^{e,}$, D.~Piccolo$^e$, G.~Saviano$^{e}$,
C.~Carrillo$^f$,
U.~Berzano$^g$, M.~Gabusi$^g$, P.~Vitulo$^g$,
M.~Kang$^h$, K.S.~Lee$^h$, S.K.~Park$^h$, S.~Shin$^h$,
A.~Sharma$^i$\\
\llap{$^a$}Ghent University, Department of Physics and Astronomy\\
Proeftuinstraat 86, 9000 Gent, Belgium\\
\llap{$^b$}Bulgarian Academy of Sciences, Inst. for Nucl. Res. and Nucl. Energy,\\
Tzarigradsko shaussee Boulevard 72, BG-1784 Sofia, Bulgaria\\
\llap{$^c$}Academy of Scientific Research and Technology of the Arab Republic of Egypt,\\
101 Sharia Kasr El-Ain, Cairo, Egypt\\
\llap{$^d$}Universita e INFN, Sezione di Bari,\\
Via Orabona 4, IT-70126 Bari, Italy\\
\llap{$^e$}INFN, Laboratori Nazionali di Frascati (LNF),\\
PO Box 13, Via Enrico Fermi 40, IT-00044 Frascati, Italy\\
\llap{$^f$}Universita e INFN, Sezione di Napoli,\\
Complesso Univ. Monte S. Angelo, Via Cintia, IT-80126 Napoli, Italy\\
\llap{$^g$}Universita e INFN, Sezione di Pavia,\\
Via Bassi 6, IT-Pavia, Italy\\
\llap{$^h$}Korea University, Department of Physics,\\
Seoul Cheongryangri 143-701, Republic of Korea\\
\llap{$^i$}CERN,\\
CH-1211 Geneva 23, Switzerland\\
E-mail: \email{michael.tytgat@cern.ch}}

\abstract{
The CMS muon system includes in both the barrel and endcap region 
Resistive Plate Chambers (RPC). They mainly serve as
  trigger detectors and also improve the reconstruction of muon parameters.
Over the years, the instantaneous luminosity of the Large Hadron Collider
gradually increases. During the LHC Phase 1 ($\sim$~first 10 years of operation)
an ultimate luminosity is expected 
above its design value of $10^{34}$~cm$^{-2}$s$^{-1}$ 
at 14 TeV. To prepare the machine and also the experiments for this,
two long shutdown periods are scheduled for 2013-2014 and 2018-2019.
The CMS Collaboration is planning several detector upgrades during these long
shutdowns. In particular, the muon detection system should be able to maintain 
a low-$p_T$ threshold for
an efficient Level-1 Muon Trigger at high particle 
rates. One of the measures to ensure this, is to extend the present RPC system
with the addition of a 4th 
layer in both endcap regions. During the first long shutdown, these two new 
stations will be equipped in 
the region $|\eta|<1.6$ with 144 High Pressure Laminate (HPL) double-gap RPCs 
operating in avalanche mode, with a similar design as the existing CMS 
endcap chambers. Here, we present 
the upgrade plans for the CMS RPC system for the fist long shutdown, 
including trigger simulation studies for the extended system, and details on 
the new HPL production, the chamber assembly and the quality control 
procedures.}


\begin{document}

\section{Introduction}

Since the start of the first Large Hadron Collider (LHC) physics run in 2009,
the Compact Muon Solenoid (CMS) experiment~\cite{cmsdetpaper} has been
collecting data successfully. During Phase 1 of the LHC which will continue 
until
about 2022, the instantaneous 
luminosity delivered to the experiments is foreseen to increase gradually up 
to more than twice its nominal value of $10^{34}$~cms$^{-2}$s$^{-1}$. At
present, two 
shutdown periods are scheduled to give the machine and the experiments 
the necessary time to
anticipate these luminosity increases: Long Shutdown 1 (LS1) in 2013/2014 
should prepare the accelerator to run at its nominal luminosity and 
Long Shutdown 2 (LS2) in
2018/2019 should take it to 2.2 times this value. 

\begin{figure}[h]
\begin{center}
\includegraphics[width=.65\textwidth]{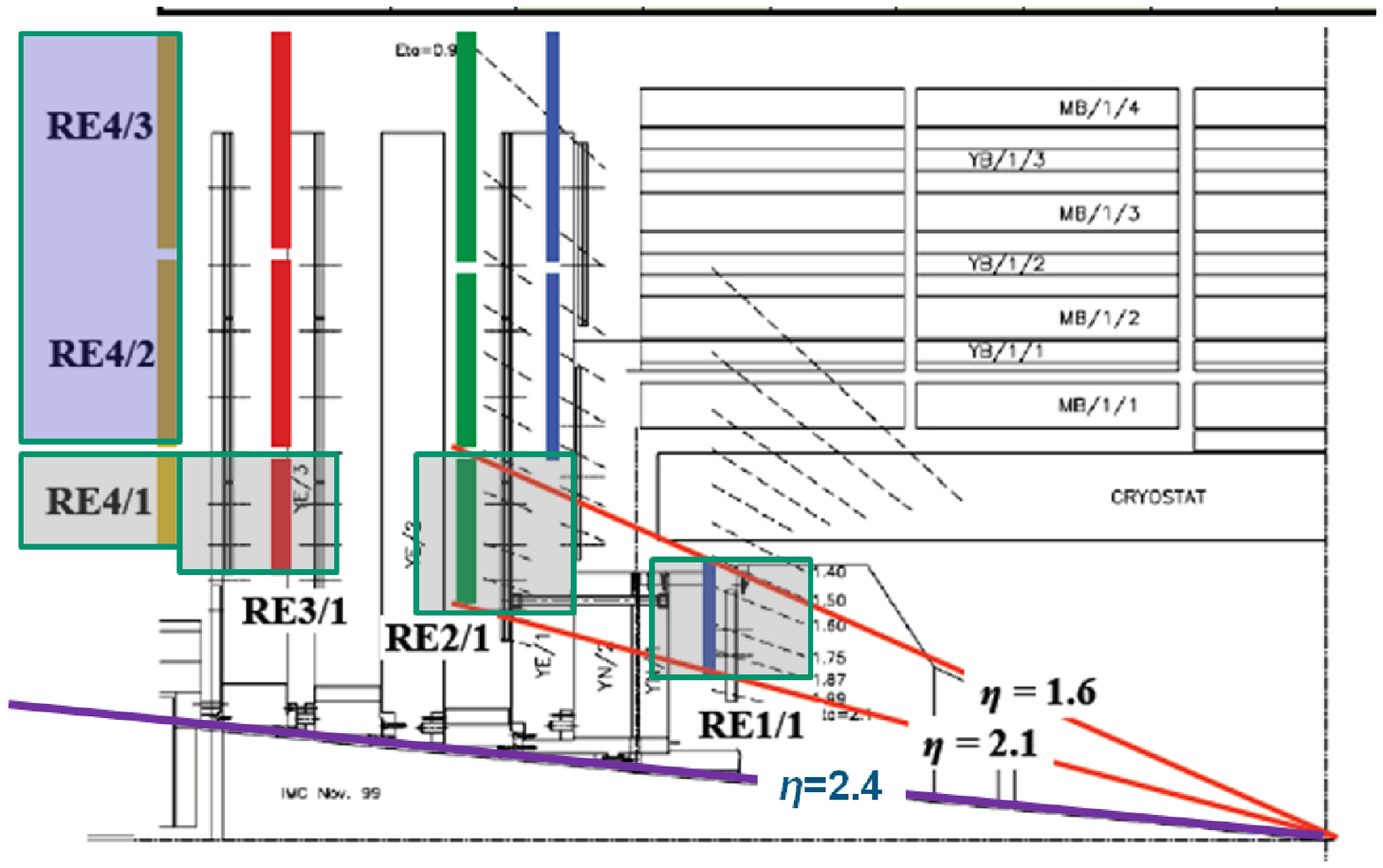}\hspace*{.03\textwidth}
\includegraphics[width=.3\textwidth]{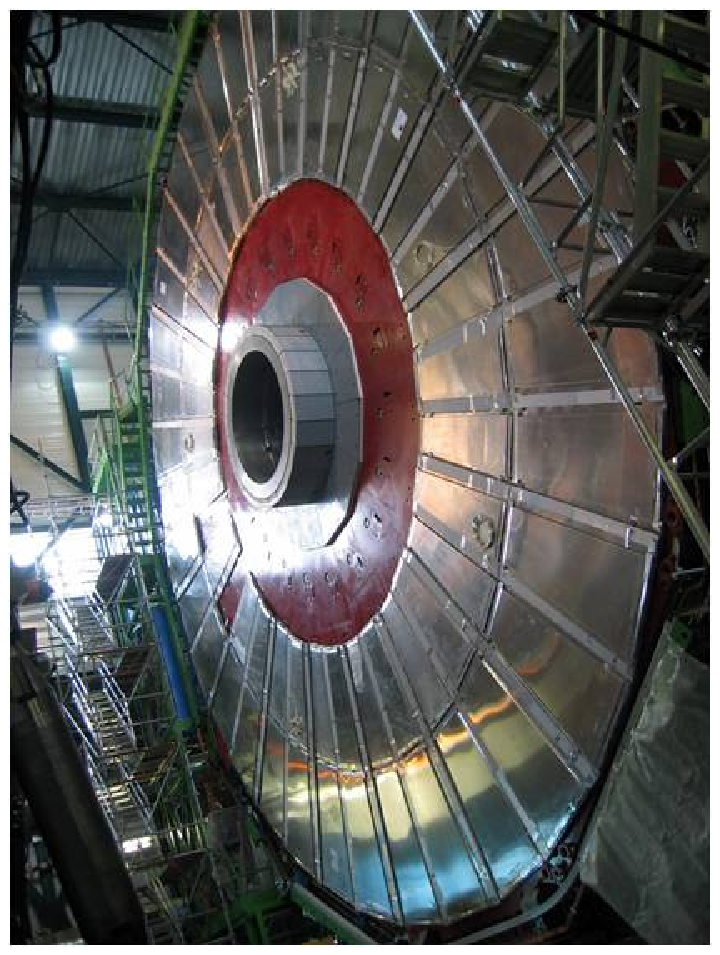}
\caption{\label{fig:cmsmuonsystem}(left) One quarter of the CMS muon
  system. The colored detectors are the RPCs in the endcap system. RPCs
  enclosed in boxes have yet to be installed. (right) Photograph of endcap RPC
  detectors taken during installation of the CMS detector.}
\end{center}
\end{figure}

During these long shutdown periods the CMS Collaboration 
intends to upgrade several subsystems of its
detector~\cite{cmsupgradeTP}. In particular, the instrumentation of
the muon system as depicted 
in Fig.~\ref{fig:cmsmuonsystem} will be
extended in both endcaps
to ensure efficient muon triggering and reconstruction in that region 
at high luminosities. In the endcaps, CMS is using Cathode Strip Chambers 
(CSCs) as muon
tracking and trigger detectors, while Resistive Plate Chambers (RPCs) serve
as dedicated trigger detectors and improve the muon reconstruction.
At present, the 4th endcap disks remain largely uninstrumented: 
CSCs are only installed in the very forward region 
($1.8<|\eta|<2.4$), and
RPCs are missing completely and cover only the first three endcap disks 
up to $|\eta|=1.6$.

During LS1 
these 4th endcap stations will be instrumented further with 
new CSCs below $|\eta|=1.8$ and new RPCs up to $|\eta|=1.6$. 
In the following, we 
describe the CMS plans for this first step in the upgrade of the RPC endcap 
system.

For the latter subsystem the very forward region beyond
$|\eta|=1.6$ will still remain empty after LS1 
and could in principle in a second step
be instrumented 
up to $|\eta|=2.4$ matching the CSC system. However, the present 
design of the endcap RPCs, made of a double High Pressure Laminate (HPL) 
gas gap and operating in
avalanche mode, is not expected to be suitable for the particle rates
amounting to several tens of kHz/cm$^2$ in the scenario of an LHC luminosity 
going up to $10^{34-35}$~cm$^{-2}$s$^{-1}$. Dedicated R\&D activities 
to identify suitable
technologies to instrument that particular endcap region during LS2 is
ongoing, see for instance~\cite{ksleerpc2012, ieee2011paper}.

\section{Motivation for the RPC Upgrade}

Muons with high transverse momenta are one of the key objects in the 
detection of possible new physics phenomena. 
By 2015, the LHC luminosity should reach $10^{34}$~cm$^{-2}$s$^{-1}$. The 
in-time pileup will be right at the edge of the CMS design envelope and will 
present special challenges for the muon system to trigger on 
high-$p_T$ muons.
The RPC upgrade is essentially driven by the impact of
the instantaneous peak luminosity on the trigger system.
In the endcaps, the RPC system 
provides excellent timing with a somewhat worse momentum resolution compared
to the CSC system. To be 
effective, the muon trigger
must achieve good enough resolution to identify high-$p_T$ tracks. 
With the RPC trigger 
requiring segments in at least three stations, the endcap system does not have 
the necessary redundancy to control the
trigger rate at the increased luminosity while preserving high trigger 
efficiency. The problem stems from mis-measurements of low-$p_T$ muons, that
are promoted to high-$p_T$ muons and contribute to the trigger rate. 
With the much higher flux of
low-momentum muons at increased luminosity, such poorly measured muons 
will dominate the trigger rate and make it unacceptably high, leading in turn
to unacceptably high muon-$p_T$ trigger thresholds. 

\begin{figure}[h]
\begin{center}
\includegraphics[width=.6\textwidth]{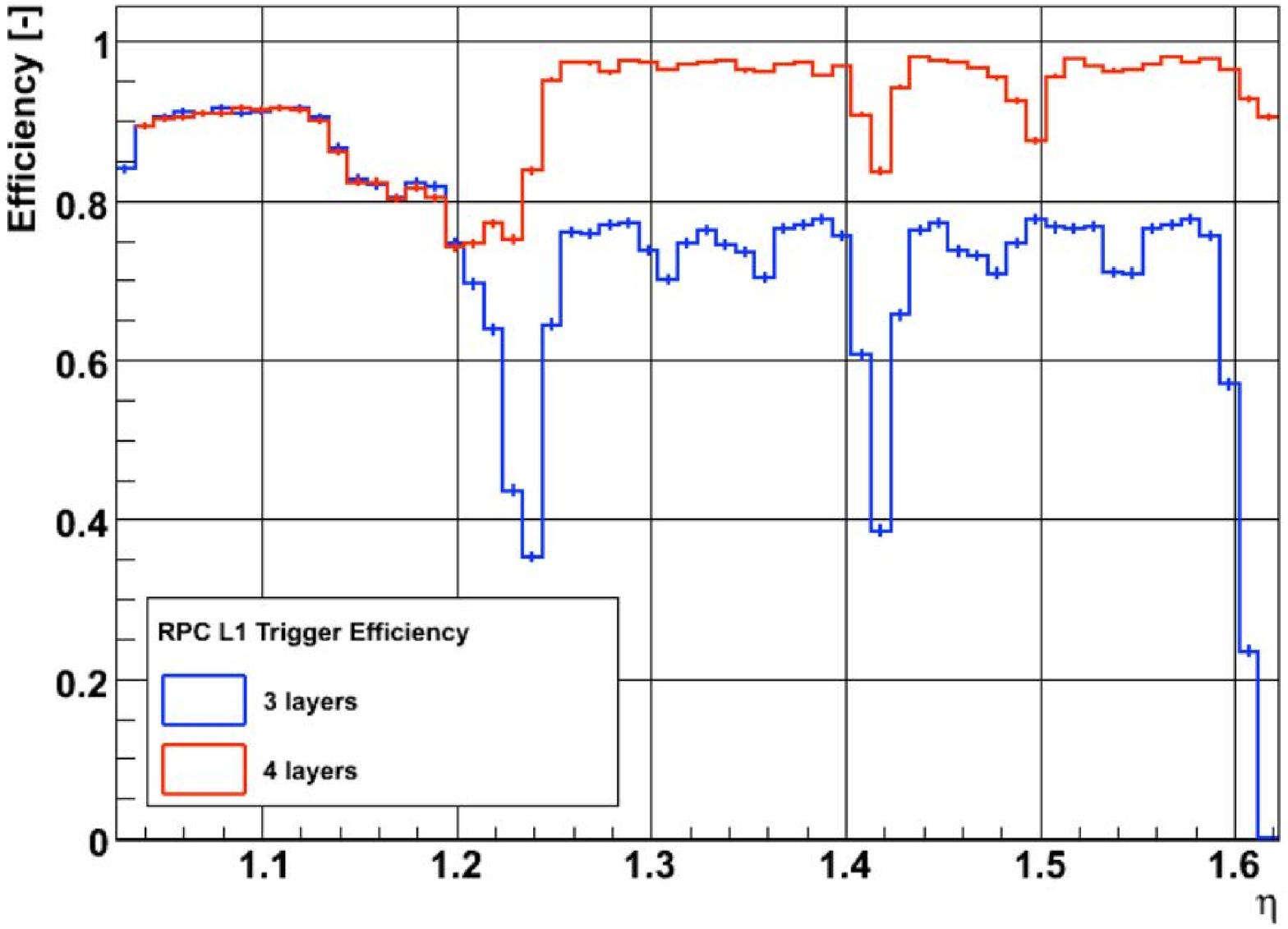}
\caption{\label{fig:RPCL1TriggerEff}Simulated RPC Level-1 trigger efficiency
  for the present system with three endcap layers compared to the upgraded
  situation with four endcap layers.}
\end{center}
\end{figure}

In Fig.~\ref{fig:RPCL1TriggerEff} the difference in the RPC Level-1 Trigger
performance is simulated between the present system with three stations and the
four station situation after the upgrade. 
The present RPC trigger logic requires hits in
at least three layers, which causes the observed drop in efficiency for the 
endcaps with only three stations. Adding the
4th layer in the endcaps, enabling a 3-out-of-4 trigger logic in
those regions, will bring the RPC endcap performance to a similar
level as the barrel system. 

\section{Design of the New CMS Endcap RPCs}

\begin{figure}[h]
\begin{center}
\includegraphics[width=.4\textwidth]{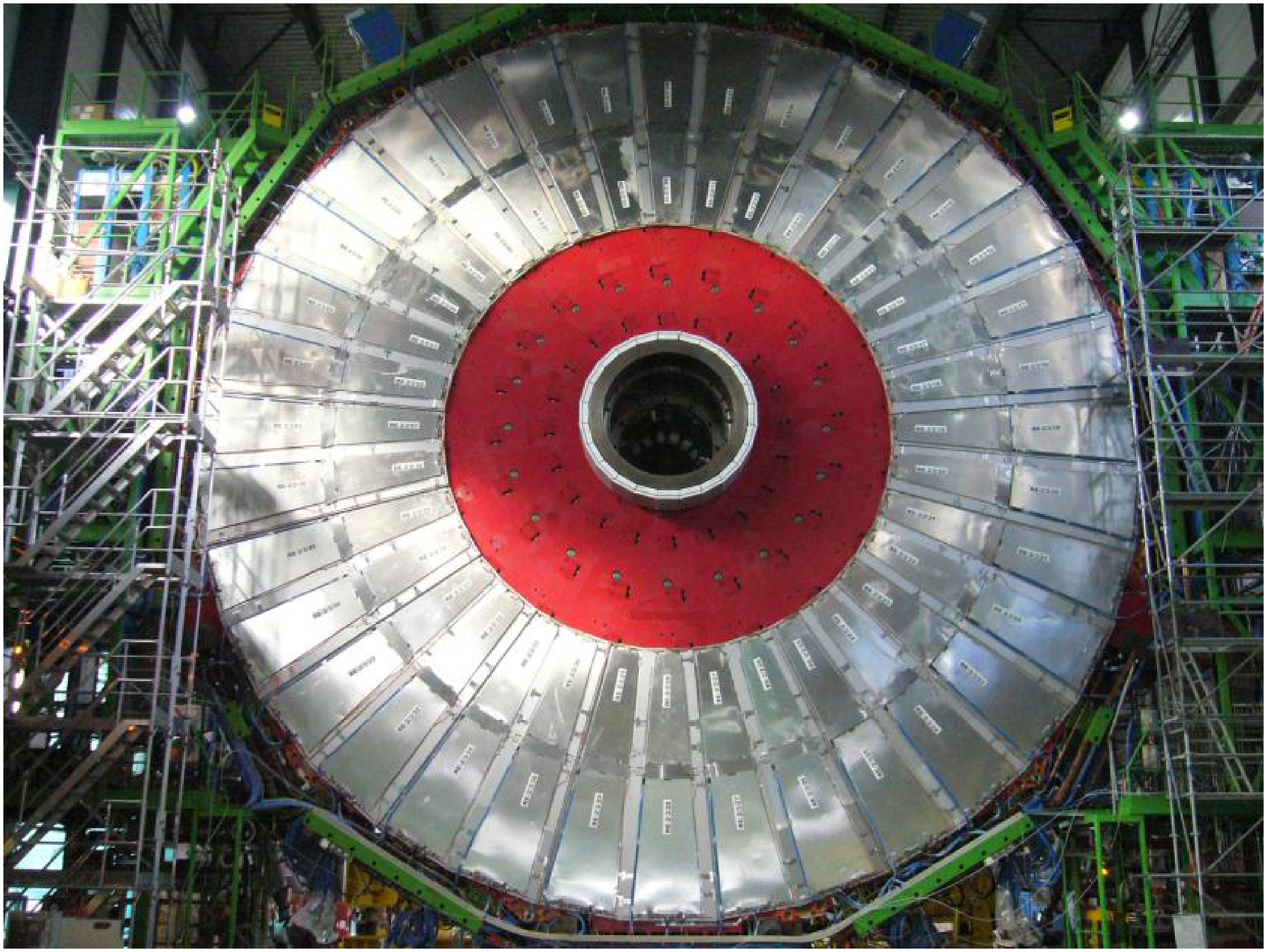}\hspace*{.03\textwidth}
\includegraphics[width=.55\textwidth]{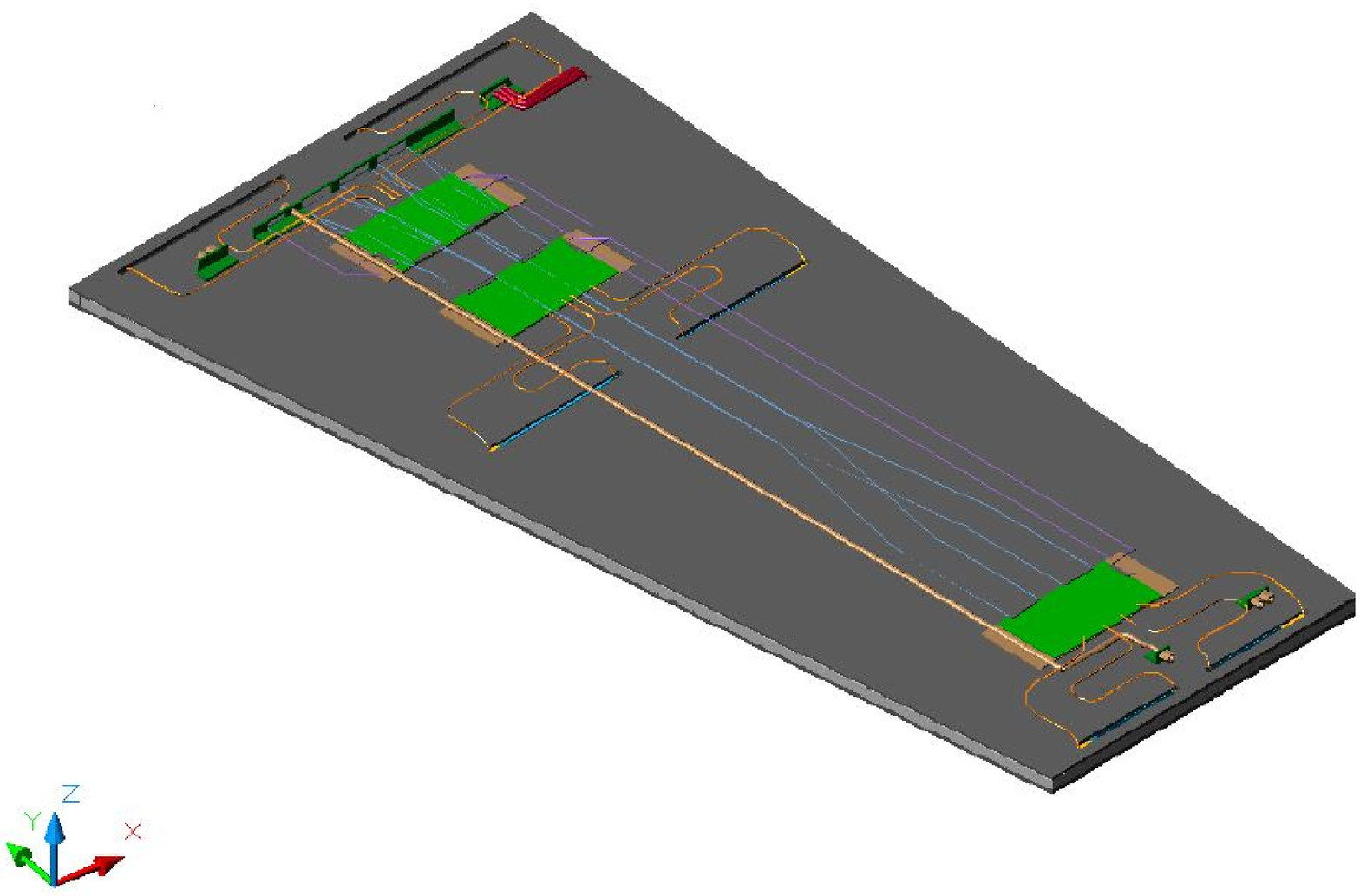}
\caption{\label{fig:layouts} (left) The second RPC endcap
  station during assembly of the CMS detector. (right)
  Schematic layout of an endcap chamber.}
\end{center}
\end{figure}

The layout of the CMS RPC endcap stations and chambers is depicted in
Fig.~\ref{fig:layouts}. Each station 
consists of three concentric rings,
called REx/1-3 (station x=1,2,3), 
with chambers mounted in a staggered way. For the 
instrumentation of the
4th station up to $|\eta|=1.6$, 144 new RPCs are required
to be mounted in two such concentric rings (RE4/2 and RE4/3) per
station, with 36 chambers per ring. As the design of the new chambers is
quite similar to the one of the existing RPCs, a total of 200 new chambers
will be built with 56 spare chambers for the RE2-4 stations.
 
The endcap RPCs contain a double HPL gas gap with a 
copper strip readout 
panel placed in between. In fact, the design uses one large gas gap at the
bottom and two smaller gaps on top of the chamber.
The chamber is embedded in a honeycomb
box with the chamber services (readout electronics, gas in and outlets and
water cooling circuit) mounted on the outside. 

The readout strip panel is divided in 3 $\eta$-partitions, with 32 strips each,
yielding a total of 96 strips per chambers. Strips are connected through coaxial
cables to Adaptor Boards linked to 3 Front End Boards (FEBs) per
chamber. Every Chamber has 1 Distribution Board for the electronics
control. The off-detector electronics consist mainly of Link Boards (LBs) that
receive signals in LVDS (Low-Voltage Differential Signaling) standard
from the FEBs and perform the synchronization with
the LHC clock and the transmission to the Trigger Electronics in the control
room. Each LB crate contains a Control Board (CB) that drives the crate,
provides inter-crate communication and takes care of the connection to the
readout and trigger systems. The LBs and CBs will include minor design
improvements related to low voltage regulation and protection circuits, to 
cure problems observed in the past with the present system.

The working point of the present system depends strongly on the
environmental temperature, and the latter should therefore be kept inside 
21-23$^{\circ}$C. To monitor the temperature, 
each endcap chamber will be equipped with one Fibre Bragg 
Grating 
sensor mounted inside a heat conducting housing; 
these sensors are radiation hard,
insensitive to magnetic fields, have a high precision, produce no 
electrical noise, are easy to install,
require a minimal amount of cabling and finally cost about the same as 
conventional 
sensors. The latter cannot be used in this case due space limitations for
cabling. 
As the new RE4 RPCs will be mounted inside the CMS detector 
facing the CSC electronics, there is some danger of additional external
heating to the RPC gas gaps. The water cooling circuit on the chambers 
consists of Cu pipes 
soldered onto three copper plates where the FEBs are mounted. An aluminum
screen box is covering the cooling circuit and the FEBs.   
To better stabilize the operating temperature, the layout and size of the
copper plates was optimized using 
FloEFD/pro\footnote{FloEFD$^{\mbox{\tiny TM}}$ software package 
by Mentor Graphics} 
and the final configuration was 
successfully tested in the laboratory with a detector inside a heated box. In
addition to that, every 10$^{\circ}$ sector of the endcap disk will get a 
connection to the
cooling water manifolds in the cavern, i.e. only 
2 chambers per connection. 

\begin{figure}[h]
\begin{center}
\includegraphics[width=.47\textwidth]{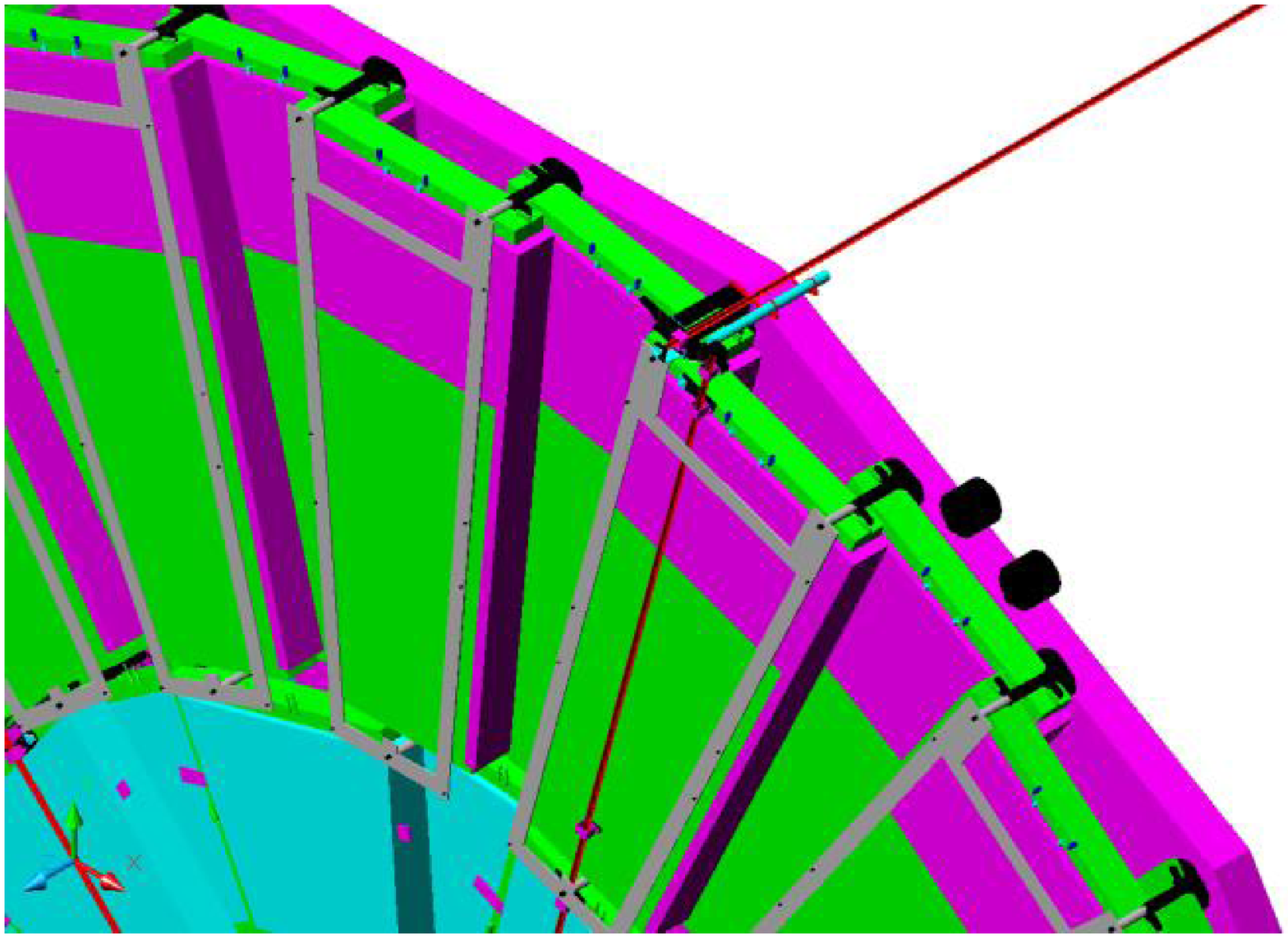}\hspace*{.03\textwidth}
\includegraphics[width=.47\textwidth]{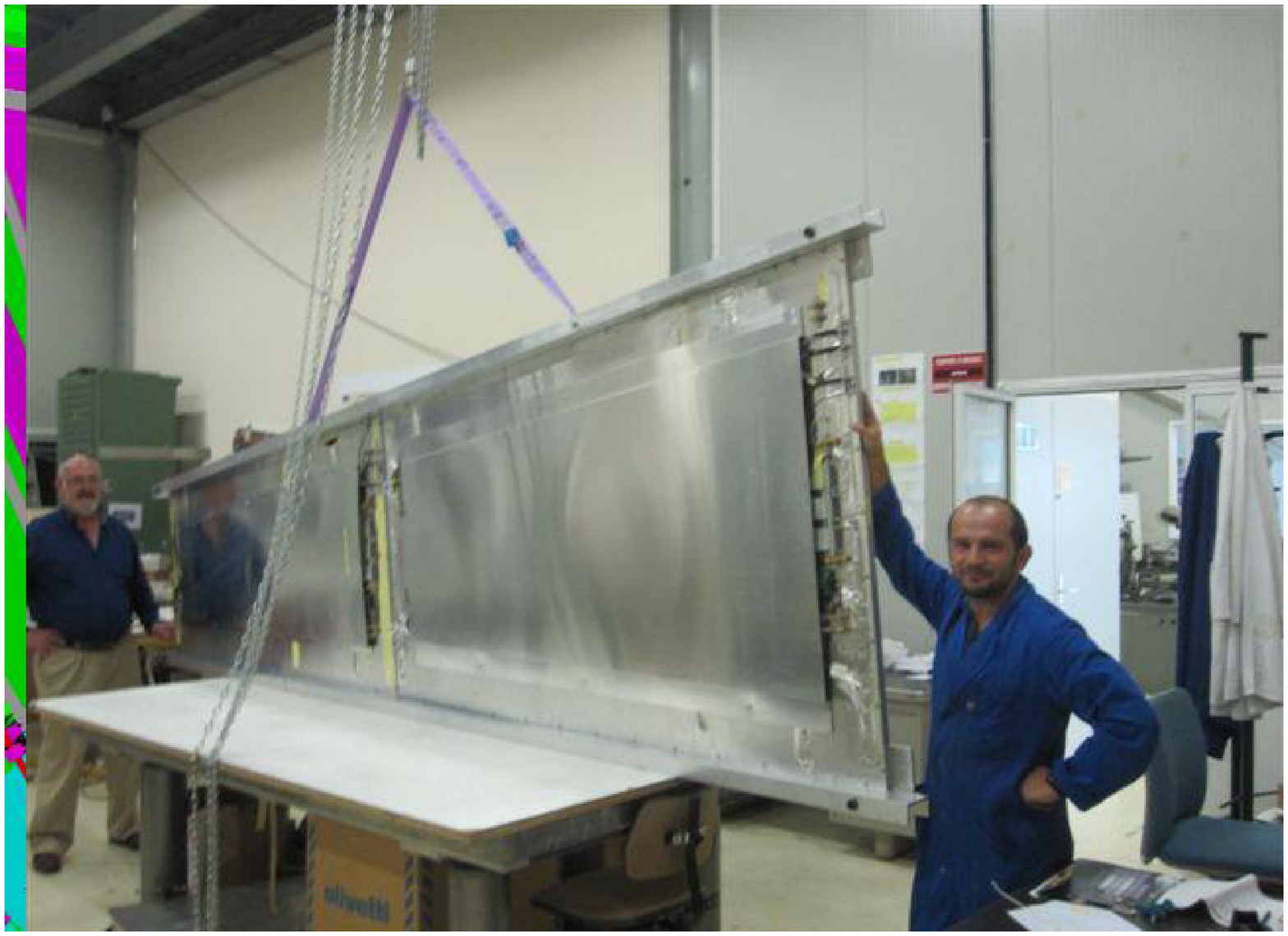}
\caption{\label{fig:RE4Installation} 
(left) Drawing of the new RPC modules installed 
  on the CSC mounting posts on the backside of the 3rd endcap yoke. (right)
  Photograph of an RE4 Super Module test installation at CERN.}
\end{center}
\end{figure}

As shown in Fig.~\ref{fig:RE4Installation}, the new RE4 RPCs will be attached 
on CSC mounting posts
on the back of the 3rd endcap yoke of the CMS detector. Before installation, 
the corresponding
RE4/2 and RE4/3 chambers will be pre-mounted together 
on Super Modules that cover a
10$^{\circ}$ sector of an endcap. This will not only reduce the amount of
cabling work to be done in the CMS cavern, but will also speed up the
installation of the chambers.

\section{RPC Production and Quality Control}

The fabrication of the RPC gas gaps starts with the
production of HPL sheets. The raw material is coming from the Puricelli
firm (Milano), where about 600 sheets of $1620\times 3200$~mm$^2$ are being 
produced. The quality control (QC) of the HPL sheets is handled by
INFN Pavia. Next, the sheets are cut by Riva (Milano) and finally surface
cleaned by General Tecnica (Frosinone).

\begin{figure}[h]
\begin{center}
\includegraphics[width=.6\textwidth]{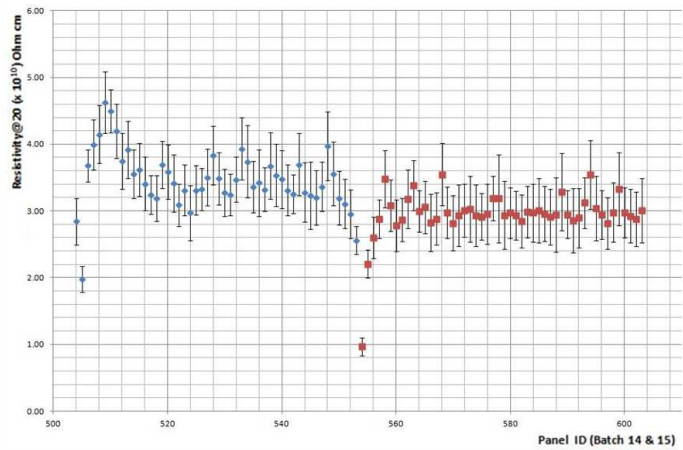}
\caption{\label{fig:bakelite}Measured resistivity for recent production
  batches of HPL sheets.}
\end{center}
\end{figure}

CMS has put forward detailed specifications for the HPL sheets; main
requirements are a high uniformity in resistivity across the surface of
the sheets and also between the sheets, with a resistivity value between 
$1-6\cdot10^{10}~\Omega$cm (measured at $20^{\circ}$C). 
Fig.~\ref{fig:bakelite} displays the resistivity values measured for two recent
production batches. The data represent the average resistivity value and spread 
measured at nine different positions across the sheet surface; 50 panels are
produced per batch and the obtained 
resistivity values are clearly uniform and well
inside the CMS specifications.

\begin{figure}[h]
\begin{center}
\includegraphics[width=.4\textwidth]{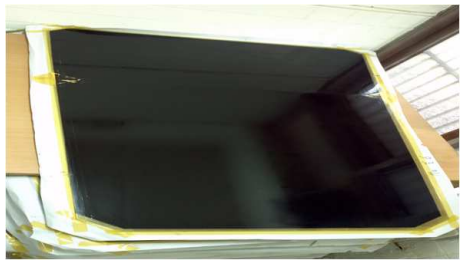}
\caption{\label{fig:gasgap}Photograph of a completed HPL gas gap at Kodel.}
\end{center}
\end{figure}

The HPL sheets that pass the QC are sent to the KOrea DEtector Laboratory 
(KODEL) at Korea University, where
about 700 gas gaps will be produced~\cite{sparkrpc2012}. The HPLs receive a
graphite coating, protected with a polyethylene terephthalate (PET) film. 
Assembled gaps are treated with
linseed oil mixed with heptane. 
Again very detailed technical
specifications and QC protocols were prepared by
CMS to ensure high quality gaps. Fig.~\ref{fig:gasgap} shows an example of a 
HPL gas gap assembled at
KODEL.    

The chamber mechanics (honeycomb boxes, screen boxes, read-out strip
panels, etc.) were produced by
two Chinese companies: Beijing Axicomb Technology Co., Ltd and Beijing
Gaonengkedi SGT Co., Ltd. In the past, those companies have already produced 
the same components for the existing system.  

The detector electronics are procured in Pakistan. For this project a total of
1350 Adaptor Boards, 650 Front End Boards and 250 Distribution Boards will be
produced. The new off-detector electronics will be produced by the INFN, while
the CMS Warsaw (Poland) group will take on the responsibility to integrate the
new electronics in the trigger system.

The final detector assembly is shared by three different institutions: 
Barc (Mumbai, India), Ghent University (Ghent, Belgium) and CERN. Each of
these assembly sites has built up a new RPC lab, including setups for detailed
quality control of the HPL gas gaps and completed chambers.  

A common quality control protocol for the chamber production has been carefully 
prepared for each level of the construction. Individual components need to
satisfy detailed mechanical specifications. The HPL gas gaps will undergo visual
inspections, leak tests and dark current measurements. Assembled chambers will
be thoroughly tested (efficiency, cluster size, long term current monitoring,
noise) in cosmic test benches at each of the assembly sites. The
tests at chamber level will be repeated once the chambers have been
transported to CERN and yet again when the chambers have been mounted in the
Super Modules. The final checkout will occur after the installation of
the Super Modules in the CMS detector. 
 
The full history of each chamber and every component is stored in an
Oracle based Construction Database. The database includes all measurements
performed before, during and after the detector assembly. This should enable
CMS to follow the evolution of each chamber in time, especially in case 
problems should appear during operation later on.

\section{Summary}

As the instantaneous luminosity of the LHC continues to increase over time
during Phase 1,
the CMS experiment has to ensure its detector can operate in a stable and
efficient manner for increasingly high particle rates. In particular, the
muon system should be able to keep its trigger rate and efficiency under 
control at a low enough muon $p_T$-threshold for physics studies. 
To this end, CMS will among other things
extend its RPC system with a 4th layer in both endcaps. 
The construction of 200 new RPCs of the standard CMS type 
recently started. Components for the new chambers are being produced at
various places in the world. The assembly and quality control of the 
chambers will be handled at three 
different locations in India, Belgium and CERN.
The new chambers are to be installed inside the CMS detector
during the first LHC Long Shutdown in 2013-2014.   

\acknowledgments

We acknowledge support from FWO (Belgium); MEYS (Bulgaria); CERN; INFN (Italy); NRF and WCU (Korea); Swiss Funding Agencies (Switzerland).

\end{document}